\documentclass[fleqn,usenatbib]{mnras}
\usepackage{newtxtext,newtxmath}
\usepackage[T1]{fontenc}

\DeclareRobustCommand{\VAN}[3]{#2}
\let\VANthebibliography\thebibliography
\def\thebibliography{\DeclareRobustCommand{\VAN}[3]{##3}\VANthebibliography}

\usepackage{graphicx}	% Including figure files
\newcommand{\swift}{\textit{Swift}}
\newcommand{\bexrb}{BeXRB}
\newcommand{\bexrbs}{BeXRBs}
\newcommand{\scnew}{Swift~J004516.6$-$734703}
\newcommand{\cs}{CSD}
\title[\scnew\ = SXP 146.6]{The newly discovered Be/X-ray binary \scnew\ in the SMC: witnessing the emergence of a circumstellar disc}

\author[J.~A. Kennea et al.]{J.~A. Kennea,$^{1}$\thanks{E-mail: jak51@psu.edu (JAK)}
M.~J. Coe,$^{2}$
P.~A. Evans$^{3}$,
I.~M. Monageng$^{4,5}$,
L.~J. Townsend$^{4}$,
\newauthor
M.~H. Siegel$^{1}$,
A. Udalski$^{6}$,
and D.~A.~H. Buckley$^{4}$
\\
$^{1}$Department of Astronomy and Astrophysics, The Pennsylvania State University, 525 Davey Lab, University Park, PA 16802, USA\\
$^{2}$Physics \& Astronomy, The University of Southampton, SO17 1BJ, UK\\
$^{3}$University of Leicester, X-ray and Observational Astronomy Research Group, School of Physics \& Astronomy, University Road, Leicester LE1 7RH, UK\\
$^{4}$South African Astronomical Observatory, P.O Box 9, Observatory, 7935, Cape Town, South Africa\\
$^{5}$Department of Astronomy, University of Cape Town, Private Bag X3, Rondebosch 7701, South Africa\\
$^{6}$Astronomical Observatory, University of Warsaw, Al. Ujazdowskie 4, 00-478 Warszawa, Poland 
}

\date{Accepted XXX. Received YYY; in original form ZZZ}

\pubyear{2020}

\begin{document}
\label{firstpage}
\pagerange{\pageref{firstpage}--\pageref{lastpage}}
\maketitle

\begin{abstract}
We report on the discovery of \scnew, a Be/X-ray binary system by the \swift\ SMC Survey, S-CUBED. \scnew, or SXP~146.6, was found to be exhibiting a bright ($\sim10^{37}$ erg/s) X-ray outburst on 2020 June 18. The historical UV and IR light-curves from OGLE and \swift/UVOT showed that after a long period of steady brightness, it experienced a significant brightening beginning around 2019 March. This IR/UV rise is likely the signature of the formation of a circumstellar disc, confirmed by the presence of strong a $H\alpha$ line in SALT spectroscopy, that was not previously present. Periodicity analysis of the OGLE data reveals a plausible 426~day binary period, and in X-ray a pulsation period of 146.6s is detected. The onset of X-ray emission from \scnew\ is likely the signature of a Type-I outburst from the first periastron passage of the neutron star companion through the newly formed circumstellar disc. We note that the formation of the circumstellar disc began at the predicted time of the previous periastron passage, suggesting its formation was spurred by tidal interaction with the neutron star. 
\end{abstract}


%%%%%%%%%%%%%%%%%%%%%%%%%%%%%%%%%%%%%%%%%%%%%%%%%%

%%%%%%%%%%%%%%%%% BODY OF PAPER %%%%%%%%%%%%%%%%%%

\section{Introduction}

Be/X-ray Binaries (\bexrbs) consist of a Be star, a massive typically main-sequence star with B spectral type showing strong Balmer and other emission lines \citep{Porter03}, and a compact object, which is most often a neutron star (NS), but sometimes a white dwarf (WD) \citep{Coe20} or a black hole \citep{Munar14}. The compact object is often in a highly eccentric orbit, with typical orbital periods of {tens to hundreds} of days \citep{Reig2011}. Notably, Be stars show excess infrared (IR) emission over regular B-stars, which is thought to be a signature of emission from a circumstellar disc (\cs) that forms around the star, although the mechanism of this formation is not clear \citep{Rivinius2013}. When the compact object approaches periastron, accretion from the \cs\ will occur, and  seen in X-rays as a short lived (days - weeks) X-ray outburst, which are referred to as Type-I outbursts. Less frequently \bexrbs\ show large, sometimes super-Eddington, outbursts (e.g. as in the recent outburst of SMC X-3, \citealt{Townsend16}), which can last for several orbital periods \citep{Reig2011}. These are referred to as Type-II bursts. When a \bexrb\ contains a NS compact object, an X-ray pulsar is often seen. 

The Small Magellanic Cloud (SMC) is an irregular dwarf galaxy, and neighbour to the Milky Way. Clearly visible by eye in the southern hemisphere, the SMC enjoys low line-of-sight absorption and a well defined distance (62.1 kpc; \citealt{Graczyk2014}), making it ideal for studies of X-ray binaries. The SMC is known to contain an over abundance of \bexrbs\ compared to the Miiky Way, and as such has been extensively surveyed in order to study and catalog \bexrb\ systems (e.g. \citealt{Haberl00,Sturm13}).

The \swift\ Small Magellanic Cloud Survey (S-CUBED; \citealt{Kennea18}) is a shallow, high cadence survey of the SMC performed by the Neil Gehrels \swift\ Observatory \citep{gehrels04}. S-CUBED consists of 142 tiled pointings covering the optical extent of the SMC (including the "Wing"). Excepting for observational and subscription constraints, S-CUBED is performed weekly, with each tile exposing for 60s. S-CUBED collects data utilizing the \swift X-ray Telescope (XRT; \citealt{burrows05}) and Ultraviolet/Optical Telescope (UVOT; \citealt{Roming05}). Given the low background of the XRT, S-CUBED is sensitive to outbursts of accreting pulsars of $>1-2\%$ Eddington luminosity. This makes S-CUBED a powerful tool for both early detection of outbursts and regular monitoring of \bexrb\ sources. 

In this letter we present the discovery, by S-CUBED, of the \bexrb\ system \scnew. We present results from both long term monitoring of this new source with S-CUBED data from the \swift\ XRT and UVOT, along with post outburst follow-up observations with \swift\ and the South African Large Telescope (SALT; \citealt{Buckley06}). In addition we present long term monitoring observations of this source by Optical Gravitational Lensing Experiment (OGLE; \citealt{Udalski2003,udalski2015}). Discussions regarding the apparent onset of Be activity, and the later onset of X-ray emission are presented.

\section{Observations}

\subsection{Discovery by S-CUBED and \swift/XRT follow-up}

\scnew\ was first discovered in outburst by the S-CUBED survey. Utilizing an automated transient pipeline set up to detect outbursting X-ray sources, we were alerted to the presence of a new X-ray source in the S-CUBED observation taken on 2020 June 18. The source was internally designated SC1774 by S-CUBED, and with the standardized \swift\ name of \scnew. 

Examination of the automatically generated X-ray light-curve of \scnew\ showed that it was detected on 2020 June 18, at a count rate of $0.289 \pm 0.075$~count s$^{-1}$, and a week earlier on 2020 June 11,  at a lower count rate of $0.037^{+0.036}_{-0.02}$ count s$^{-1}$, showing almost an order of magnitude brightening. Although S-CUBED began on 2016 June 8, no previous X-ray detection of the source had been made by the survey.

After discovery we requested a rapid turn around \swift\ Target-of-Opportunity Request (TOO) for 2~ks monitoring observations with a 2 day cadence with XRT in Photon Counting (PC) mode in order to obtain a high quality spectrum and short term light-curve. UVOT was configured to obtain images in the 6 optical and UV filters. These observations commenced on 2020 June 19 {and continued until 2020 August 16, with a total of 24 TOO observations taken. A gap in monitoring between the observations taken on 2020 June 27 and 2020 July 7 was caused by an observing constraint.} These data were analysed utilizing standard XRT analysis tools provided by HEAsoft 6.27.2 and using the methods described by \cite{Evans09}.

We calculated a X-ray position for \scnew, utilizing the method described by \cite{Goad07}, using UVOT data to correct for the systematic errors in astrometry. We find a position of RA(J2000) $= 00^h\ 45^m\ 17.91^s$, Dec(J2000) = $-73^\circ\ 47'\ 05.5''$, with an error radius of $1.9''$ (90\% confidence). This position does not match any previously known X-ray source, but is consistent with an optical source, 2dFS0556, with a spectral type of B1-3(III) \citep{Evans04}. Analysis of the UVOT data finds a counterpart at the position RA(J2000) = $00^h\ 45^m\ 17.77^s$, Dec(J2000) = $-73^\circ\ 47'\ 05.591''$, which lies $0.6''$ from the center of the XRT error region, and $1''$ from 2dFS0556. 

In the first TOO observation the source was well detected at an average brightness of $0.387 \pm 0.021$ count s$^{-1}$ in XRT data. The PC mode X-ray spectrum is well described by an absorbed power-law model, with $N_H$ fixed at a standard value of $5.9\times10^{20}\ \mathrm{cm}^{-2}$ \citep{dl1990}, we find a photon index $\Gamma = 0.50\pm 0.16$ with a reduced $\chi^2 = 0.89$ (14 degrees of freedom). We note that the hard X-ray spectrum is consistent with other High Mass X-ray Binaries (HMXB) and \bexrbs\ in the SMC. Given the companion is a B-star, the onset of X-ray emission this is highly suggestive that \scnew\ is a newly discovered \bexrb\ system.

The fitted peak flux, corrected for absorption is $3.0 \pm 0.4 \times 10^{-11}\ \mathrm{erg \ cm^{-2}\ s{^-1}}$ (0.5 $-$ 10~keV). Assuming a standard SMC distance of 62.1 kpc, this corresponds to a 0.5 $-$ 10 keV luminosity of $1.5 \times 10^{37}\ \mathrm{erg\ s^{-1}}$.

\begin{figure}
	\begin{center}
	\includegraphics[width=8.5cm]{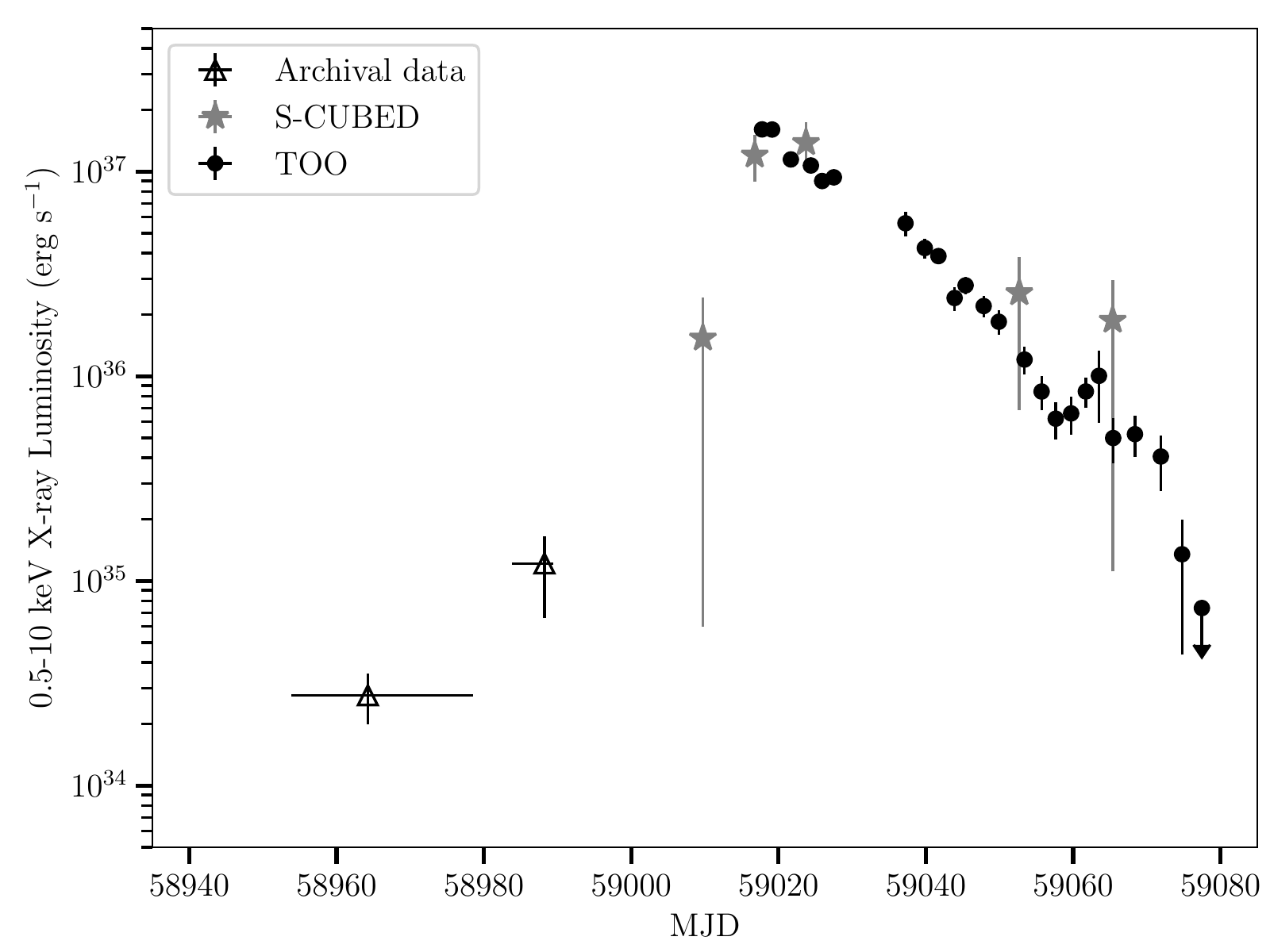}
		\end{center}
    \caption{X-ray light-curve of combined archival, S-CUBED and TOO data from the \swift/XRT for \scnew, during the first X-ray outburst seen from this \bexrb. In the final observation taken on 2020 August 16 it is not detected, so an upper limit is shown.}
    \label{fig:xray_lc}
\end{figure}

It was noted that this new transient lies near ($2.4'$) the recently discovered candidate Be/WD X-ray Binary system, Swift~J004427.3$-$734801 \citep{Coe20}, therefore archival \swift/XRT data from detailed monitoring of this source will contain pre-outburst observations of \scnew. We analyzed these observations and found that the source was detected in these observations between 2020 April 14, and 2020 May 20, albeit at a much fainter level. Combining the data from the 25 observations of the source over this time period, with a combined exposure time of 38.4~ks, with the TOO and S-CUBED observations, we created a light-curve of \scnew\ (Fig.~\ref{fig:xray_lc}). This light-curve shows that the luminosity (assuming that the spectrum has not varied) ranged between $3.6 - 6.7 \times 10^{34}\ \mathrm{erg \ cm^{-2}}$ (0.5 - 10 keV) before the S-CUBED detection, and showed a steady rise before the first detection by S-CUBED. The combined X-ray light-curve in Fig.\ref{fig:xray_lc} is consistent with the rise, peak and fall of a \bexrb\ Type-I outburst.

\begin{figure}
	\begin{center}
	\includegraphics[width=8.5cm]{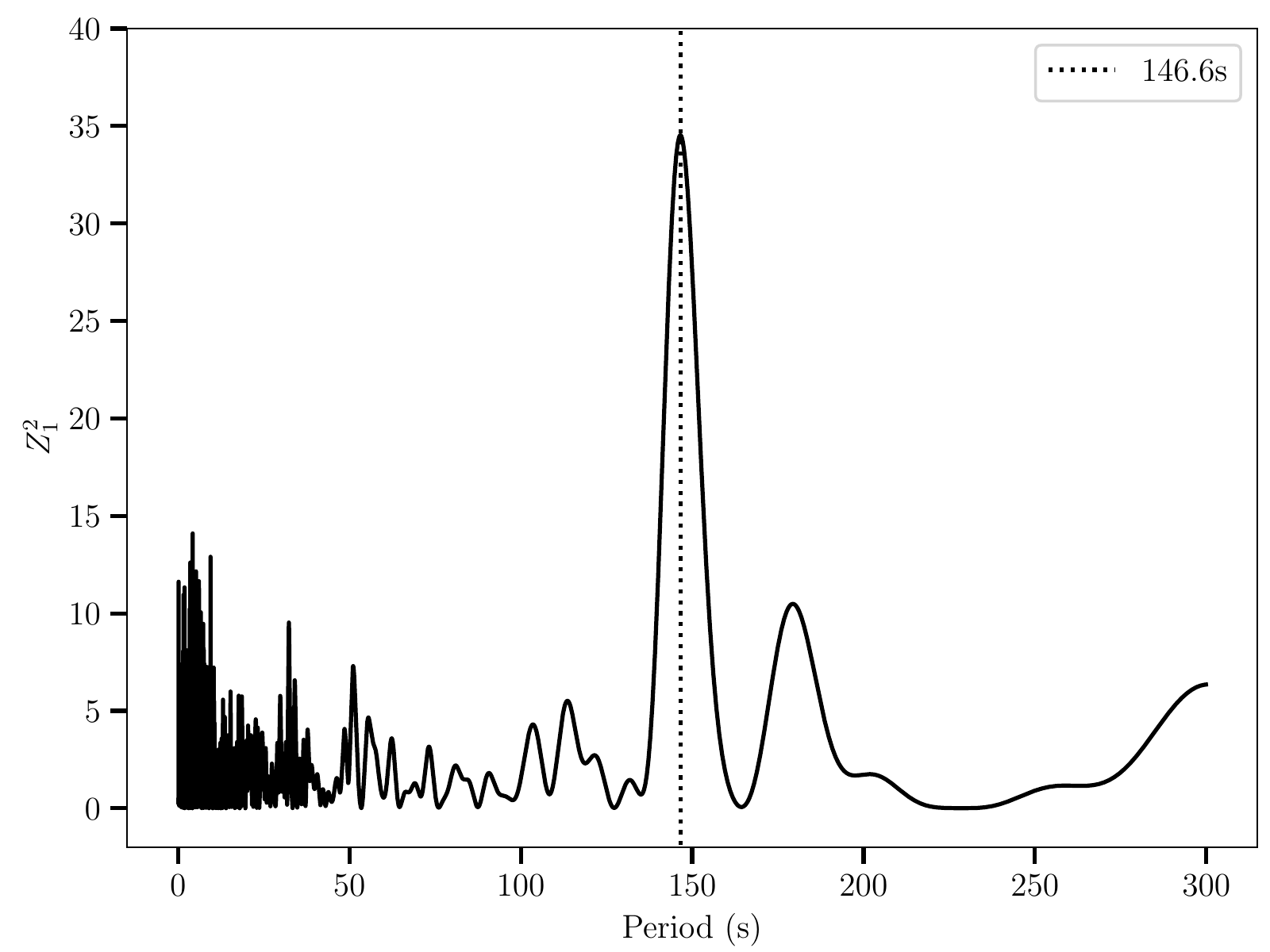}
		\end{center}
    \caption{$Z^2_1$ periodogram of \swift/XRT data of \scnew\ taken on 2020 June 21, showing a clear peak at $P=146.6$~s (marked with dotted line). All observations of \scnew\ at peak show similar detections, albeit at lower significance due to few total counts.}
    \label{fig:pulsar_periodogram}
\end{figure}

We searched the post-outburst TOO observations for the presence of pulsations using a $Z^2_1$ search \citep{Buccheri83}, and find a significant detection of a pulsar period of $P = 146.6\pm1.1$~s (error estimated using the Monte-Carlo method of \citealt{Gotthelf99}), detected in the observation of 2020 June 21{, shown in Fig.~\ref{fig:pulsar_periodogram}}. The pulsation is detected, albeit with a lower significance in the observations of 2020 June 17 and 2020 June 19. Utilizing the nomenclature defined by \cite{Coe05}, we suggest an alternative name for \scnew\ of SXP 146.6.

Combining all S-CUBED data taken before 2020, we do not detect the source, and calculate a $3\sigma$ upper limit on the average count rate of $2.1 \times 10^{-3}\ \mathrm{count\ s^{-1}}$, which assuming the spectral fit above, is the equivalent to an upper limit on the pre-outburst luminosity of $<8 \times 10^{34}\ \mathrm{erg\ s^{-1}}$ ($0.5-10$ keV).

\subsection{Optical, IR and UV observations}

\subsection{OGLE data}

\begin{figure}
	\includegraphics[width=8.5cm]{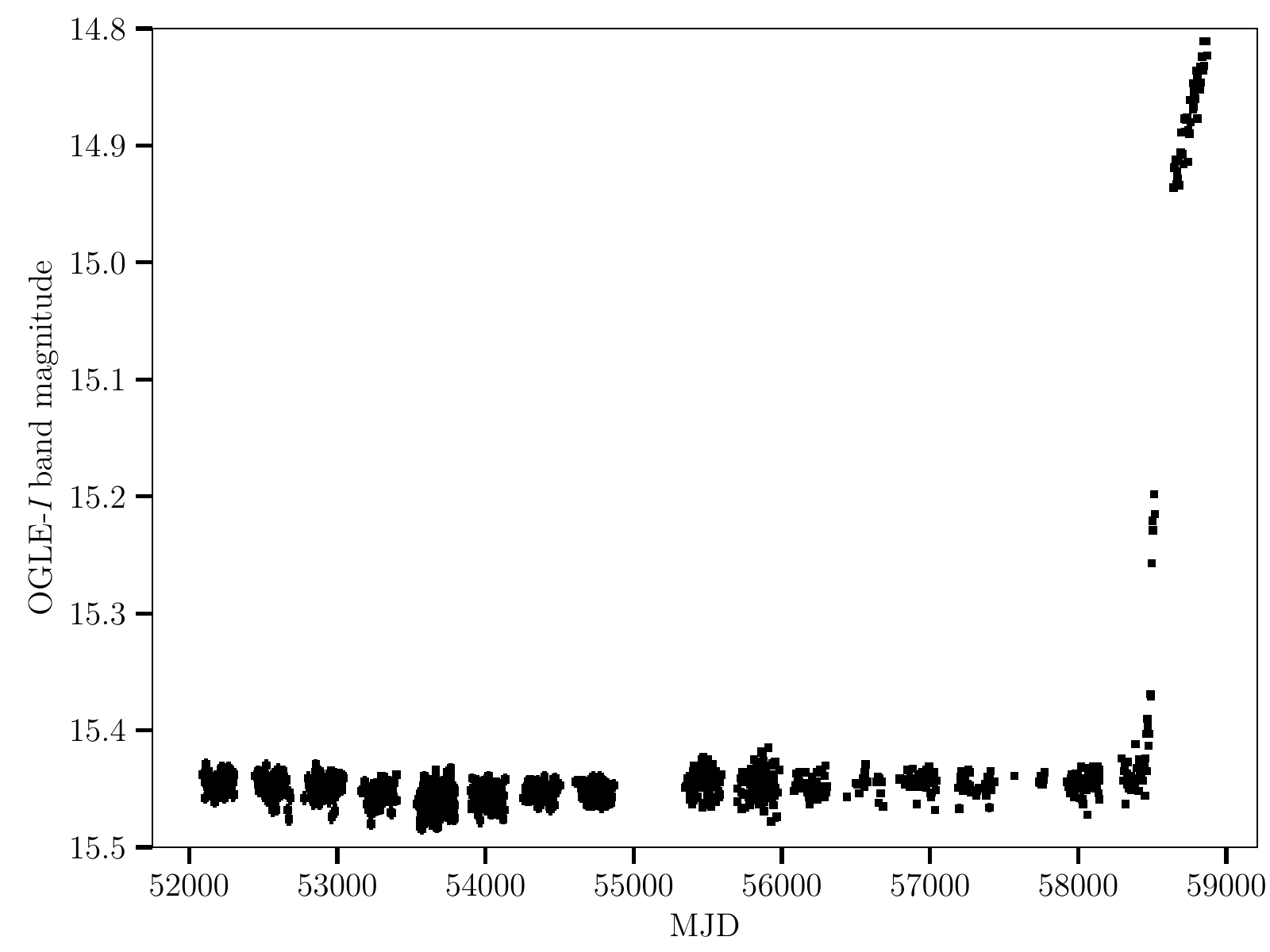}
    \caption{Long term 20-year light-curve of \scnew\ from OGLE.}
    \label{fig:ogle}
\end{figure}

The OGLE project provides long term \textit{I}-band photometry with a cadence of 1-3 days. it was possible to retrieve many years of photometric monitoring from OGLE III \& IV in the \textit{I}-band. The optical counterpart to \scnew\ was identified with the OGLE source smc128.3.12493 in OGLE III and smc720.12.13583 in OGLE IV. As a result it is possible to construct a 20 year light-curve in the \textit{I}-band - see Fig.~\ref{fig:ogle}. This striking light-curve reveals a source that has been in a low state for at least 17-18 years with no evidence of flaring activity throughout this time. Then around MJD 58500 (2019 January 16) it began to brighten and continued this dramatic change through to the end of the current OGLE data coverage, with no sign that this increase was levelling off.

\begin{figure}
	\begin{center}
	\includegraphics[width=8.5cm]{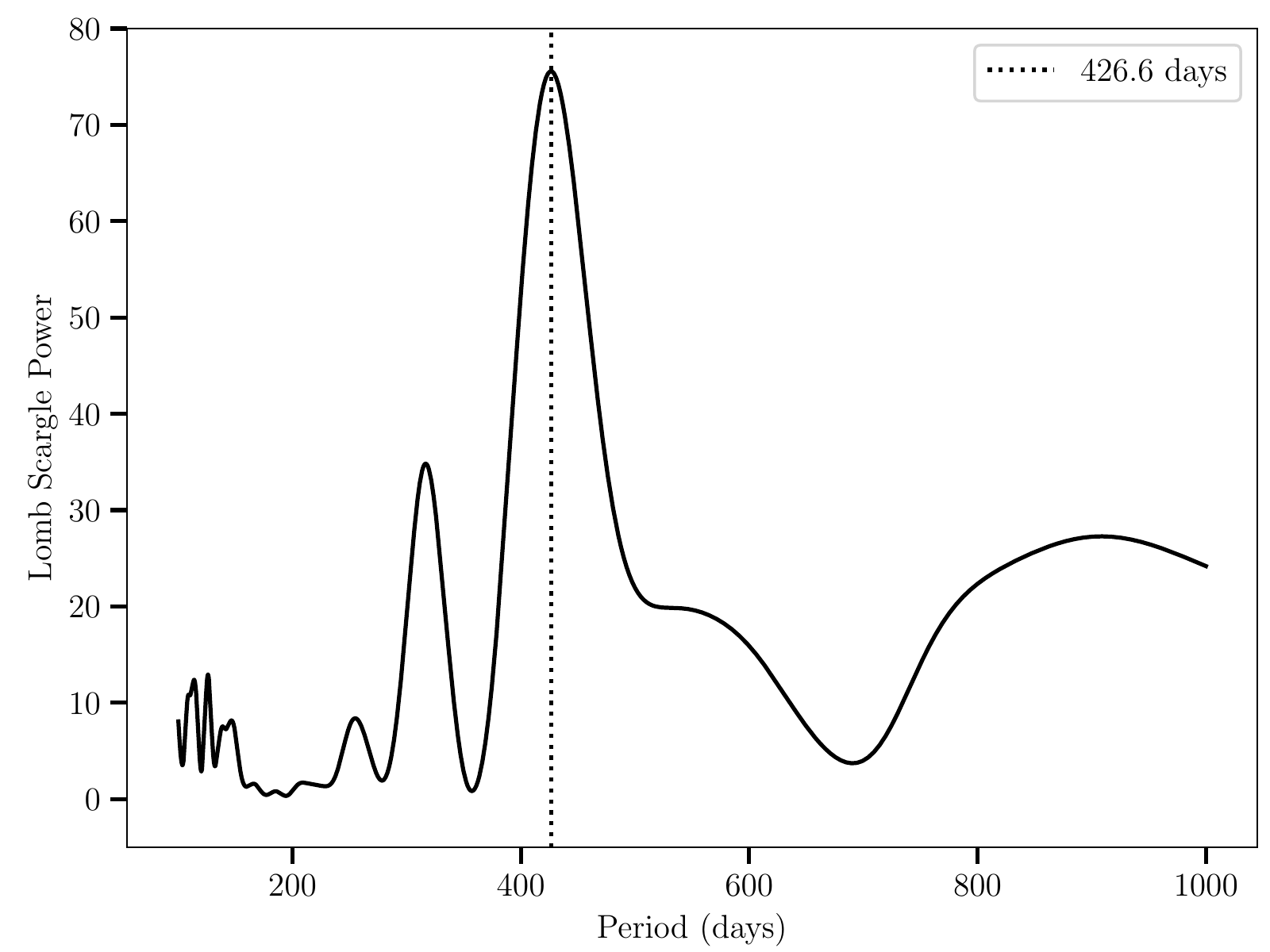}
		\end{center}
    \caption{Lomb-Scargle periodogram of the OGLE-\textit{I} band data showing a strong peak at $426.4 \pm 1.8$~days, marked with a dotted line, likely to be the orbital period of \scnew. }
    \label{fig:ogle_periodogram}
\end{figure}

\begin{figure}
	\includegraphics[width=8cm]{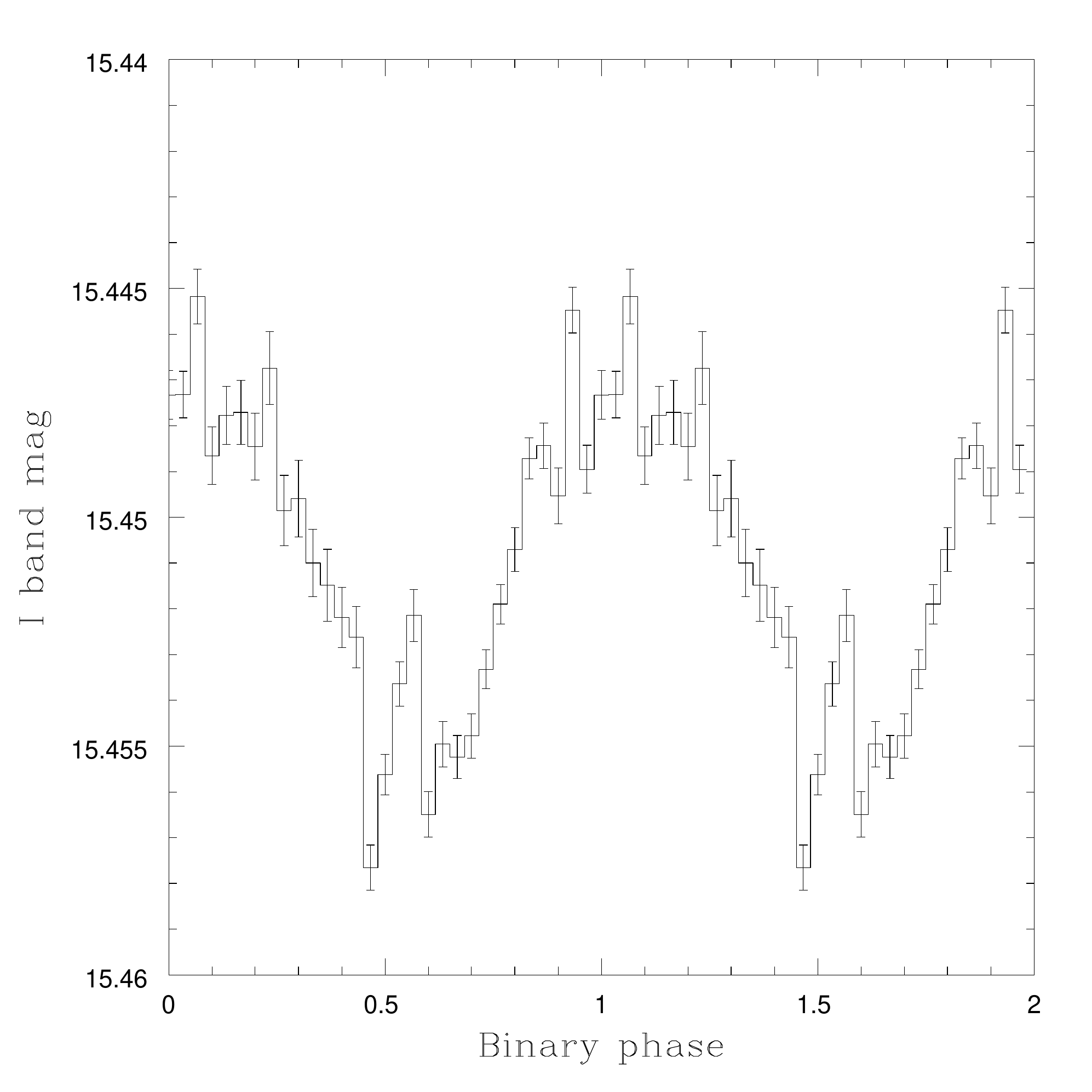}
    \caption{All OGLE data prior to the current outburst folded at a period of 426d.}
    \label{fig:fold}
\end{figure}

If the data prior to the start of the outburst are analysed for possible periodicities using the Lomb-Scargle technique then a very strong peak emerges in the power spectrum at a period of 426.4 $\pm$1.8 days {(see Fig~\ref{fig:ogle_periodogram})}. Because of the 18 year length of this data run it is easy to distinguish this peak from the smaller, annual one rising from the data sampling. If the data are folded at this period then a sinusoidal profile is revealed with an amplitude of $\sim$0.005 magnitudes - see Fig.~\ref{fig:fold}. This small modulation is only detectable because the multi-year length of the data run.

It seems very probable that this is the binary period of the system with the NS partner inducing small regular changes in the structure of outer envelope of the star, or in a minimal \cs. Using the profile shown in Fig.~\ref{fig:fold} it is possible to determine an ephemeris for the brightest part of the 426d cycle of:

\begin{equation}
    T_\mathrm{peak} = n(426.4\pm1.8) + 52133.5\ \mathrm{MJD}
    \label{eq:ephem}
\end{equation}

It is likely that this brightest phase corresponds to the closest approach of the NS each orbit, though there may be some phase delay in the response of the B-type star and the brightest time may not be precisely at periastron.

\subsection{\swift/UVOT data}

\begin{figure}
	\includegraphics[width=8cm]{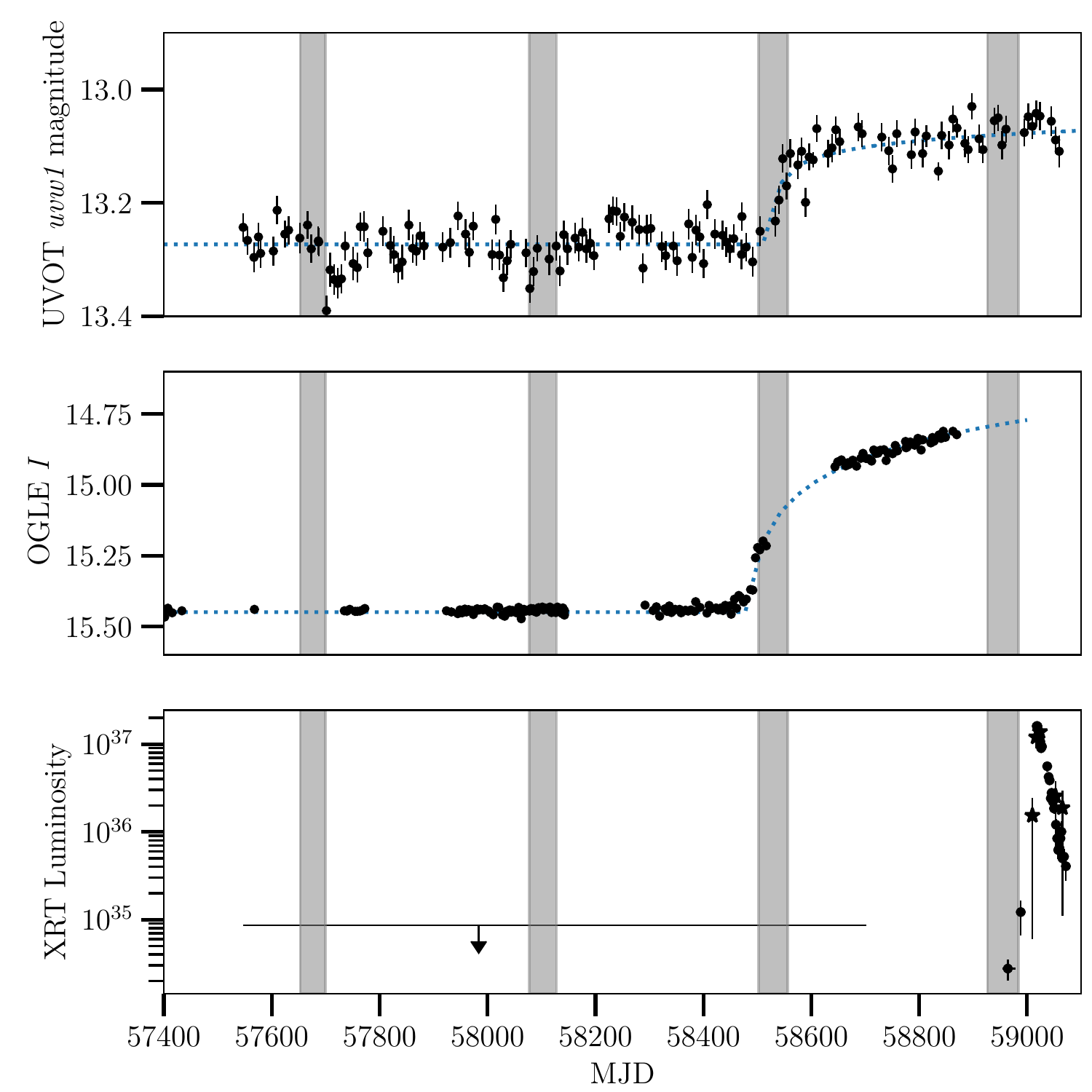}
    \caption{Light-curves of \scnew\  on the timescale of the entire S-CUBED survey. Top panel, UVOT \textit{uvw1} filter data from S-CUBED and other observations, middle panel shows OGLE \textit{I}-band data, bottom panel shows S-CUBED X-ray luminosity points. Blue dotted lines show a simple parametric model fit to OGLE and UVOT data. Grey bars show the predicted time of optical peaks based on the OGLE derived ephemeris given in Equation\ref{eq:ephem}, the width of the grey bar indicates the uncertainty. }
    \label{fig:light-curves}
\end{figure}

UVOT observations of \scnew\ have been taken by S-CUBED since it began in June 2016, approximately weekly with 60s exposures utilizing the \textit{uvw1} filter.  Analysis of UVOT data were performed utilizing the \texttt{uvotmaghist} tool from HEAsoft v6.27.2. We have extracted the \textit{uvw1} light-curve for \scnew\ for the whole S-CUBED dataset. The optical counterpart is well detected in all UVOT data, with a brightness $\mathit{uvw1} = 12.8 - 13.2$. The resultant light-curve is shown in Fig.~\ref{fig:light-curves}.

The light-curve is notable in that for the period from the start of S-CUBED monitoring (2016 June 8) until approximately 2019 Feb 28, the \textit{uvw1} light-curve is essentially flat, with $\mathit{uvw1} = 13.27 \pm 0.01$ (statistical error only). After this period it shows a steady rise finally reaching 
$\mathit{uvw1} = 13.07\pm0.02$, 
over approximately 200 days. This change in brightness began more than a year before the onset of the X-ray brightening. Examination of the recent UVOT data show no significant variability in any of the 6 UVOT filters, despite the rise in X-rays. The mean UVOT brightness for all TOO observations  are given in Table~\ref{tab:uvot_magnitudes}. 

\begin{table}
	\centering
	\caption{Mean UVOT magnitudes for all combined data taken in 2020 for \scnew\ during the Type-I X-ray outburst. Note that there is no significant variability or brightening trends seen during this period.}
	\label{tab:uvot_magnitudes}
	\begin{tabular}{llcr} % four columns, alignment for each
	\hline
	\textit{u}& 13.619&$\pm$&0.004\\
    \textit{b}& 14.593&$\pm$&0.004\\
    \textit{v}& 14.561&$\pm$&0.007\\
    \textit{uvw1}& 12.959&$\pm$&0.004\\
    \textit{uvw2}& 12.747&$\pm$&0.003\\
    \textit{uvm2}& 12.839&$\pm$&0.004\\
	\hline
	\end{tabular}
\end{table}

\subsection{SALT data}

\begin{figure*}
    \centering
    \includegraphics[width=18cm]{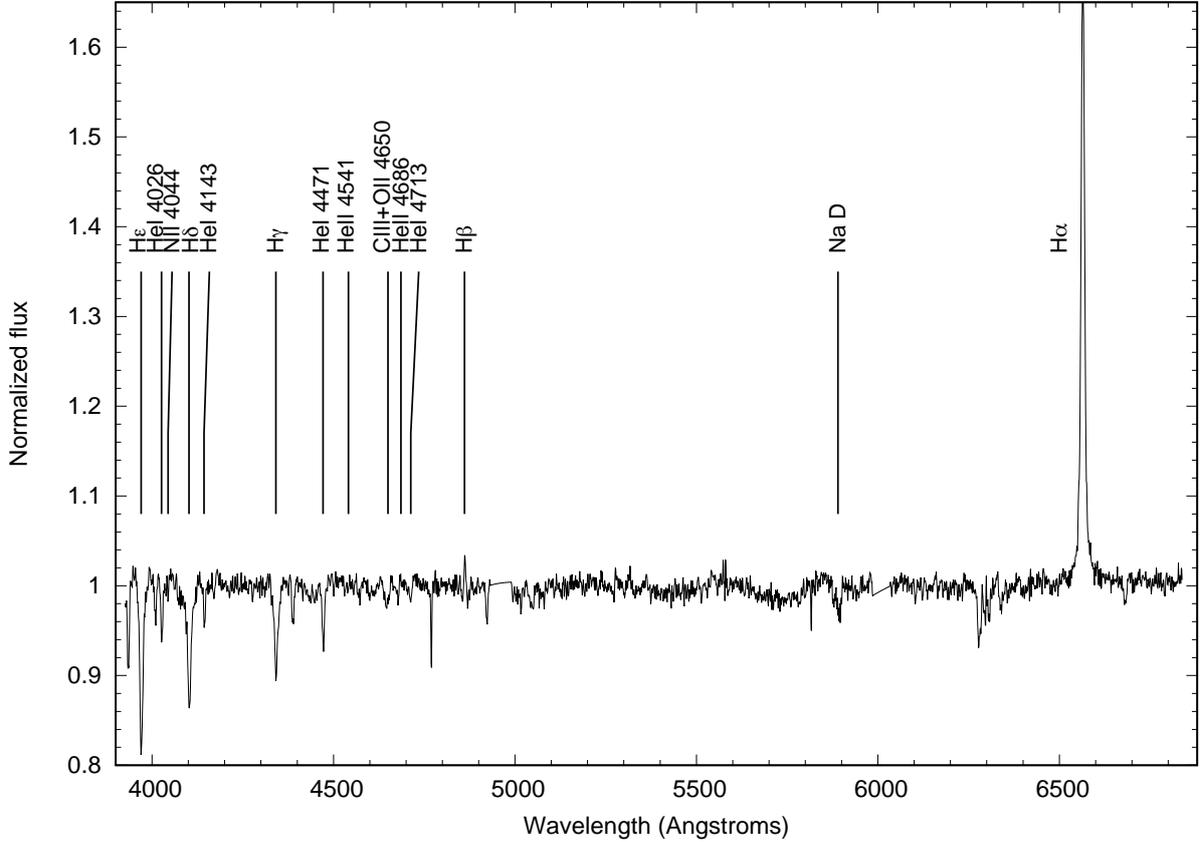}
    \caption{Normalised SALT spectrum of \scnew. The different line species are labeled at their expected rest wavelengths.}
    \label{fig:salt_spec}
\end{figure*}

A single broadband optical spectrum of \scnew\ was obtained with SALT. The observation was executed early in the morning on 2020 19 June (MJD 59019) using the Robert Stobie Spectrograph (RSS; \citealt{Burgh03}) in long slit mode. The PG0900 grating was used to obtain a spectrum between 3950--7000 \AA, at a resolution of approximately 1700 at H$_{\alpha}$ and 1100 at H$_{\gamma}$. The initial reductions (overscan and gain correction, bias subtraction, and amplifier cross-talk correction) were performed using the SALT pipeline \citep{Crawford12}. Wavelength calibration, background subtraction and the extraction of the 1D spectrum were done using various tasks in \textsc{iraf}\footnote{Image Reduction and Analysis Facility: iraf.noao.edu}. Fig.~\ref{fig:salt_spec} shows the resulting spectrum with all the relevant line species shown.

The SALT spectrum shows a strong, single peaked emission line at H$_{\alpha}$ immediately confirming the suspected Be nature of the counterpart. A measured equivalent width of -10.2 \AA\ is moderate given the scale of the optical and X-ray outbursts, 
though this is not atypical for other \bexrbs. Archival observations \citep{Evans04} show that all Balmer lines were previously in absorption, indicating the sudden appearance of a \cs\ is the most likely cause of the current X-ray outburst. H$_{\beta}$ is also slightly in emission, though an accurate equivalent width is difficult to obtain given the weakness of the line.

\section{Discussion}

\subsection{The spectral class of the B star}

The signal in the blue end is good enough to identify several helium and oxygen lines in absorption, in addition to H$_{\gamma}$ and H$_{\delta}$. The CIII+OII blend at 4650 \AA\ constrains the spectral type to earlier than B3 and the apparent absence of any He II further constrains the spectral type to later than B0.5 — though we note that the SNR is not good enough to conclude the absence of He II for certain. This range of spectral type is in broad agreement with the observations of \cite{Evans04} who quote B1-3 III. Si IV, if present, is certainly weaker than OII, making the most likely spectral type around B1.5. As a majority of metal lines in the spectrum are not so obvious, we are limited to using the HeI 4121 to HeI 4143 ratio to determine the luminosity class \citep{Walborn90}. This line ratio strengthens towards more luminous stars, suggesting that the star has a luminosity class of IV–V. Using the observed magnitude (V$\sim$15) together with the distance modulus of the SMC (18.95, \citealt{Graczyk2014}) to constrain the luminosity class also supports this classification. In summary we classify this star as B1.5 IV-V.

\subsection{Comparison with other Be/X-ray binaries}

It is interesting to compare \scnew\ with another SMC Be/X-ray binary system, SXP756 \citep{coe2004}. This system also exhibits a clear, long-term OGLE modulation at a period of 394d, similar to that proposed here for \scnew,  though the shape of the modulation is very much more prominent. In SXP756 a sharp spike in the \textit{I}-band is revealed once per orbit rather than the much broader modulation seen here. This difference is probably attributable to a much more eccentric orbit in SXP 756 and suggests that of \scnew\ may be a low eccentricity. Of course, the inclination of the orbital plane to that of the \cs\ will also be important in defining the way in which the disc reacts to the motion of the NS . Detailed modelling beyond the scope of this discovery paper is required.

\cite{Corbet1986} has shown that there is a correlation between orbital period and pulsar period, although there is a large scatter in this relationship. Given the long, 426~day orbital period, the 146.6~s pulsar period is plausible within this paradigm. 

{We note that \cite{Reig2011} characterize Type-I outbursts as having a peak X-ray luminosity around $10^{37}$~erg/s and outbursts that last approximately $0.2-0.3 P_{orb}$. The properties of the outburst seen from \scnew\ are consistent with these parameters, and are aligned with the predicted periastron passage time based on the derived orbital period from OGLE data, suggesting that we are indeed seeing a Type-I outburst, rather than a Type-II.}

\subsection{Formation of the circumstellar disc} 

The observation of Balmer lines in the optical spectrum of \scnew, the fact that in previous observations these were not present, along with the rise in \textit{I}-band magnitude which began in early 2019, after many years of steady emission, point strongly towards the recent appearance of a \cs. 

The recent onset of X-ray emission from this source is likely a Type-I \bexrb\ outburst, which was caused by the interaction of the NS with the newly formed \cs. The peak X-ray luminosity of the source is consistent with the brightness of Type-I outbursts which are in the range of $10^{36} - 10^{37}\ \mathrm{erg\ s^{-1}}$. Examining the S-CUBED X-ray light-curve, no previous outburst activity was seen. The predicted periastron passage using Eq.~\ref{eq:ephem} aligns well with the onset of the Type-I outburst. Additionally the previous periastron passage occurred during the period in which the OGLE-\textit{I} and UVOT-\textit{uvw1} light-curves were starting to rise, suggesting that although a \cs\ was beginning to form during this time, there was no significant accretion on the NS at periastron, likely as the disc had not grown to sufficient size to interact with the NS . 

In order to determine when the \cs\ began to form, we fit a simple parametric model to the OGLE and UVOT data defined as:
\begin{equation}
    \Delta m = A \log(1+B(t-t_0)^2) 
\end{equation}
where $t_0$ is the start of the outburst, $\Delta m$ is the change in brightness in magnitudes and $A$ and $B$ are constants. Note for $t<t_0$ we fixed $\Delta m = 0$.
Utilizing this fit we are able to estimate the value of $t_0$, i.e. when the IR/UV brightening began. For the OGLE-\textit{I} band data we find that $t_0 = 58481.3 \pm 1.1$~MJD. Fitting the same model to the UVOT data, we find that $t_0 =  58532.8 \pm 1.5$~MJD, $\sim51$~days after the \textit{I}-band. The reason for this lag in UV rise time is not clear, although it is suggestive that the UV emission is likely not simply the UV tail of emission from the \cs, and may be the signature of heating in the disc as it grows. % Speculation :D

The timing of apparent appearance of the \cs\ is interesting, coming as it does close to the predicted previous periastron passage to the Type-I outburst. This suggests the birth of the \cs\ could be spurred by tidal interaction of the Be star with the NS companion, which has previously been suggested as a factor in \cs\ formation \citep{Rivinius2013}.

\section{Conclusions}

The S-CUBED survey, a weekly survey of the SMC in X-ray and UV performed by \swift, discovered a previously unknown X-ray transient in the SMC, \scnew, with a pulsar period of $146.6$~s. Examination of historical and catalogued data on this star revealed that it has a B-type optical companion, and post-outburst observations with SALT revealed characteristic Balmer and other line emission in the spectrum, conclusively showing that this system was a \bexrb. However, previous observations of the star did not reveal the presence of Balmer emission lines, which along with the fact that no X-ray Type-I bursts had been seen from this source before, suggest that a \cs\ had recently formed. Examination of the long term ($\sim20$ year) light-curve from OGLE revealed that in early 2019, after a long period of inactivity, the I band flux had started to rise. Strengthening IR emission being a signature of the presence of a \cs, this suggested that we were witnessing the birth of a \cs, and that the X-ray outburst seen by S-CUBED represented the first periastron passage of the NS though the disc since it became fully formed. Detection of a plausible long (426 day) periodicity in the OGLE light-curve further supports that this recent X-ray outburst would have been the first passage since the disc formed, and tantalisingly, the formation of the \cs\ is coincident to the previous periastron passage, suggesting a possible link between the onset of \cs\ formation and tidal interactions in the binary system.

Finally, we note that the combination of the binary period and the pulse period fits comfortably on the Corbet diagram \citep{Corbet1986} confirming this empirical relationship out to the longest known binary periods.

\section*{Acknowledgements}

The OGLE project has received funding from the National Science Centre, Poland, grant MAESTRO 2014/14/A/ST9/00121 to AU. JAK acknowledges support from NASA Grant NAS5-00136. PAE acknowledges UKSA support. IMM, LJT and DAHB are supported by the South African National Research Foundation. Some of the observations reported in this paper were obtained with the Southern African Large Telescope (SALT) under programme 2018-2-LSP-001. The Polish participation in SALT is funded by grant no. MNiSW DIR/WK/2016/07.

\section*{Data Availability}

The data underlying this article will be shared on reasonable request to the corresponding author.

%%%%%%%%%%%%%%%%%%%%%%%%%%%%%%%%%%%%%%%%%%%%%%%%%%



\bibliographystyle{mnras}
\bibliography{references} 
%%%%%%%%%%%%%%%%%%%%%%%%%%%%%%%%%%%%%%%%%%%%%%%%%%

% Don't change these lines
\bsp	% typesetting comment
\label{lastpage}
\end{document}